%% file: main.tex
\documentclass[runningheads]{llncs}

\input{packages}
\input{preamble}

\begin{document}

\title{\tempo: Reconstructing Synchronous Reactive Programming with \ocaml{} 5 Effects}
\titlerunning{\tempo: Synchronous Reactive Programming with \ocaml{} 5 Effects}

\author{Fr\'ed\'eric Dabrowski}
\authorrunning{F. Dabrowski}
\institute{Univ. Orl\'eans, INSA Centre Val de Loire, LIFO UR 4022, Orl\'eans, France\\
\email{frederic.dabrowski@univ-orleans.fr}}

\maketitle

\begin{abstract}
Synchronous reactive programming gives reactive systems a deterministic
temporal structure by organizing execution into logical instants and
signal-based communication.
Boussinot's synchronous reactive model extends this setting with cooperative
threads, broadcast signals, and dynamic processes; \reactiveml{} brings that
model into a strict, typed, higher-order functional language.
This paper studies whether the same core mechanisms can be reconstructed inside
ordinary \ocaml{}~5, rather than exposed by a dedicated language extension.
We present \tempo{}, a library runtime based on algebraic effects and deep
handlers: effect operations delimit reactive suspension points, and the handler
reifies captured continuations as tasks scheduled by logical-instant semantics.
A comparative study with \reactiveml{} quantifies the overhead of this
library-level reconstruction and identifies the runtime mechanisms that
dominate its cost.

\end{abstract}

\section{Introduction}
\input{sections/introduction}

\section{Background and Problem Statement}
\input{sections/background}

\section{The Tempo Language}
\label{sec:language}
\input{sections/language}

\section{Implementation}
\label{sec:implementation}
\input{sections/implementation}

\section{Evaluation}
\input{sections/evaluation}

\section{Related Work}
\input{sections/related}

\section{Conclusion}
\input{sections/conclusion}

\bibliographystyle{splncs04}
\bibliography{references}

\end{document}

%% file: packages.tex
\usepackage[T1]{fontenc}
\usepackage{graphicx}
\usepackage{amsmath}
\usepackage{amssymb}
\usepackage{booktabs}
\usepackage{listings}
\usepackage{xcolor}
\usepackage[hidelinks]{hyperref}

%% file: preamble.tex
\newif\ifanonymous
\anonymousfalse
\ifdefined\ANONYMOUSSUBMISSION
  \anonymoustrue
\fi

\hypersetup{
  bookmarksopen=true,
  bookmarksnumbered=true,
  bookmarksdepth=1
}

\lstdefinelanguage{OCaml}{
  keywords={let,in,match,with,fun,function,type,module,if,then,else,rec},
  sensitive=true,
  comment=[n]{(*}{*)},
  morestring=[b]"
}

\makeatletter
\renewcommand\paragraph{\@startsection{paragraph}{4}{\z@}%
                       {-8\p@ \@plus -3\p@ \@minus -3\p@}%
                       {-0.5em \@plus -0.22em \@minus -0.1em}%
                       {\normalfont\normalsize\itshape}}
\makeatother

\newcommand{\code}[1]{\texttt{#1}}
\newcommand{\tempo}{Tempo}
\newcommand{\reactiveml}{ReactiveML}
\newcommand{\ocaml}{OCaml}
\newcommand{\githubtempo}{\textit{The implementation link is omitted for double-blind review.}}

\newcommand{\emitop}{\code{emit}}
\newcommand{\awaitimmop}{\code{await\_immediate}}
\newcommand{\awaitop}{\code{await}}
\newcommand{\pauseop}{\code{pause}}
\newcommand{\parallelop}{\code{parallel}}
\newcommand{\whenop}{\code{when\_}}
\newcommand{\watchop}{\code{watch}}

\newcommand{\bOne}{B1}
\newcommand{\bTwo}{B2}
\newcommand{\bThree}{B3}
\newcommand{\bFour}{B4}
\newcommand{\bFive}{B5}


%% file: sections/introduction.tex
Reactive systems maintain an ongoing interaction with their environment:
they react to events, update internal state, and coordinate concurrent
activities over potentially unbounded executions.
Synchronous programming gives this coordination a deterministic temporal
structure by organizing execution into logical instants
~\cite{berry1992esterel,caspi1987lustre,halbwachs1991lustre,benveniste2003synchronous}.
Boussinot's synchronous reactive model adapts this discipline to cooperative
threads and broadcast communication: absence is not observed while an instant
may still produce new signals, but only once same-instant propagation has
stabilized
~\cite{boussinot2002reactive,boussinot2006fairthreads}.
\reactiveml{} then showed that this model could be integrated into a strict,
typed, higher-order functional language
~\cite{mandel2005reactiveml,mandel2006phd}, while related designs such as Lucid
Synchrone illustrate another functional route to synchronous programming
~\cite{pouzet2006lucid}.

These languages provide a strong semantic basis for reactive programming, but
they also place the reactive boundary in a dedicated source language, compiler,
or runtime system.  \ocaml{}~5 changes the implementation landscape by making
algebraic effects and handlers available in the host language.  An effect
operation can suspend a computation, expose its continuation to a handler, and
resume it under explicit scheduler control
~\cite{plotkin2013effects,dolan2020ocaml,dolan2017multicore,leijen2017koka}.
This is the form of control needed by synchronous reactive execution:
programs suspend at temporal and signal boundaries, while the runtime decides
whether the captured continuation belongs to the current instant, a later
instant, a blocked guard, or a preemption context.

This paper studies whether the reactive boundary can be moved from a dedicated
language extension to an \ocaml{} library while preserving the synchronous
structure of the model.  To investigate this question, we introduce
\emph{\tempo{}}, a reactive programming library implemented in \ocaml{}~5.
\tempo{} provides direct-style reactive operators and interprets them through a
deep effect handler whose captured continuations become the runtime tasks of a
synchronous reactive machine.  The implementation reconstructs logical
instants, signal propagation, guarded activation, parallel composition, and weak
preemption at library level.
\githubtempo

\paragraph{Contributions.}
This paper makes three contributions.
First, it defines an effect-handler runtime architecture that reconstructs core
synchronous/reactive mechanisms without dedicated language extensions.
Second, it provides an operational account of this architecture through runtime
state compartments, signal frontiers, handler protocols, and continuation
ownership rules.
Third, it evaluates the approach both on mechanism-focused benchmarks and on
a composed supervision scenario.
The scope of these claims is intentionally empirical: results are architectural
evidence for this implementation and workload family, not a claim of
compiler-neutral performance parity or full feature equivalence with
\reactiveml{}.

The paper is organized as follows: Section~\ref{sec:background-problem}
establishes background and problem framing, Section~\ref{sec:language}
presents the \tempo{} language at user level, Section~\ref{sec:implementation}
details the runtime architecture, Section~\ref{sec:eval} reports the empirical
evaluation, Section~\ref{sec:related} discusses related work, and
Section~\ref{sec:conclusion} concludes.

%% file: sections/background.tex
\label{sec:background-problem}
The construction of \tempo{} depends on two bodies of work: synchronous
reactive programming, which supplies the temporal semantics, and \ocaml{}~5
effect handlers, which supply the runtime control mechanism.  This section
recalls the parts used later in the paper: logical instants and absence,
Boussinot's delayed treatment of absence, the higher-order functional
realization provided by \reactiveml{}, and the deep-handler mechanism used by
\tempo{}.

\subsection{Synchronous Logical Time and Absence}

The synchronous programming paradigm emerged in the 1980s to give deterministic
models of concurrency for control-oriented systems.  Languages such as Esterel,
Lustre, and Signal established its main concepts
~\cite{berry1992esterel,halbwachs1991lustre,benveniste2003synchronous}.
Their central abstraction is \emph{logical time}: execution is organized as a
sequence of discrete instants, and computations within one instant are treated
as simultaneous with respect to this logical clock.  Concurrent activities can
therefore interact through signals without exposing an arbitrary thread
interleaving.  During an instant, signals may be emitted and observed, and
control decisions may depend on their presence or absence; at the end of the
instant, the system moves to the next logical step and signal presence is reset.
This simple temporal discipline enables formal verification
~\cite{halbwachs1992programming,benveniste2003synchronous}
and compilation into efficient synchronous hardware implementations
~\cite{berry1992esterel,potop2007compiling}.
Its main semantic difficulty is the treatment of absence.  If a program can
test immediately whether a signal is absent while the instant is still open,
circular causal dependencies may arise: a signal may be emitted precisely when
it is absent.  Such causality problems, central in the theory of synchronous
languages~\cite{berry1992esterel,benveniste2003synchronous}, made absence one
of the defining issues for the reactive models that followed.

\subsection{Boussinot's Synchronous Reactive Model}

Boussinot's work develops a programming-language-oriented branch of this
tradition.  It keeps logical instants but adopts a different compromise on
absence: instead of strengthening the static causality discipline of
Esterel-like languages, absence is delayed until the instant can no longer make
progress
~\cite{boussinot2002reactive,boussinot2006fairthreads}.  While the instant is
open, only positive information is propagated.  Signals emitted during the
instant are broadcast to active computations, which may emit further signals,
create dynamic processes, or suspend.  The instant closes only when this
same-instant propagation has reached a fixpoint; at that point, signals not
emitted during the instant can be considered absent, and the corresponding
reactions are prepared for the following instant.  This discipline supports
cooperative threads, dynamic process creation, and broadcast communication.  It
also gives a programming discipline for absence that avoids many circular
causality problems without requiring the full constructive causality analysis of
Esterel.

This line of work also led to several implementations.  Before FairThreads,
Boussinot and collaborators developed Reactive-C, Reactive Scripts, and
SugarCubes in C, scripting, and Java settings
~\cite{boussinot1991reactivec,boussinot1996reactivescripts,boussinot1998sugarcubes}.
FairThreads later realized the model with cooperative execution and
events~\cite{boussinot2006fairthreads}.

\subsection{\reactiveml{} in a Higher-Order Functional Setting}

\reactiveml{} belongs directly to this lineage, but its distinctive
contribution is to place the model in a strict, typed, higher-order functional
language.  Rather than proposing another low-level runtime for the reactive
model, it extends ML with a process layer for synchronous reactive computation
~\cite{mandel2005reactiveml,mandel2006phd,mandel2015tenyears}.  The language
separates ordinary ML computations from reactive processes and gives the
temporal part of a program dedicated constructs for logical instants,
synchronous parallel composition, signal emission and observation, suspension,
dynamic process creation, and weak preemption.  This design showed that
Boussinot's relaxed treatment of absence can coexist with higher-order
abstraction, strong typing, and structured reactive control.

For \tempo{}, \reactiveml{} is both a semantic reference and a practical
comparison point.  It identifies the synchronous reactive mechanisms that
matter in an ML setting, but realizes them through a dedicated source language,
compiler, and runtime scheduler~\cite{mandel2015tenyears}.  \tempo{} keeps the
same family of mechanisms while moving the reactive boundary into ordinary
\ocaml{}, as a library-level runtime.

\subsection{Algebraic Effects and Deep Handlers}

Algebraic effects provide the control mechanism used by \tempo{} to implement
synchronous reactive execution inside ordinary \ocaml{}.  They expose
disciplined suspension and resumption in the host language
~\cite{plotkin2013effects,dolan2017multicore,dolan2020ocaml,leijen2017koka},
and connect to the broader tradition in which control constructs are
interpreted through explicit continuation manipulation
~\cite{danvy1990control,filinski1994representing}.  An effect operation is a
named request performed by a computation.  Performing it does not execute the
operation directly; it suspends the computation at the nearest enclosing
handler.  The handler receives both the operation and a continuation for the
rest of the suspended computation, which it may resume immediately, store for
later, or discard.

\ocaml{} 5 exposes both shallow and deep effect handlers~\cite{ocaml54effects}.
\tempo{} relies on deep handlers: when a handler resumes a continuation, later
effects performed by that continuation are intercepted by the same handler.
Thus the scheduler remains the control boundary for a reactive computation even
after it has been suspended and resumed.  With a shallow handler, this boundary
would have to be reinstalled explicitly around each resumed continuation.

Listing~\ref{lst:effects-cooperative} recalls a cooperative-thread example often
used to introduce handlers
~\cite{ocaml54effects,sivaramakrishnan2015effective,dolan2017multicore}.
The surface operations \code{yield} and \code{spawn} are ordinary \ocaml{}
functions; the policy appears only at the handler boundary:

\begin{lstlisting}[float=t,caption={Cooperative scheduling with effects.},
label={lst:effects-cooperative},basicstyle=\ttfamily\footnotesize,frame=none]
type _ Effect.t +=
  | Yield : unit Effect.t
  | Spawn : (unit -> unit) -> unit Effect.t
let yield () = perform Yield
let spawn f = perform (Spawn f)
let run main =
  let q = Queue.create () in
  let rec step f =
    match f () with
    | () -> next ()
    | effect Yield, k -> Queue.add (continue k) q; next ()
    | effect (Spawn f), k -> Queue.add f q; continue k ()
  and next () =
    if not (Queue.is_empty q) then step (Queue.take q)
  in step main
let example () =
  spawn (fun () -> print_endline "child");
  print_endline "parent";
  yield ();
  print_endline "parent again"
\end{lstlisting}
On \code{Yield}, the handler captures \(k\), enqueues its resumption, and runs
another thread.  On \code{Spawn(f)}, it enqueues \code{f} and resumes the parent
continuation.  Running \code{example} prints \code{parent}, then \code{child},
then \code{parent again}; a different policy for \code{Spawn} or \code{Yield}
could reorder these messages.

This dependence on a scheduler policy is precisely what synchronous reactive
semantics avoids exposing at the reactive level.  Within an instant, signal
presence is determined by logical-time propagation rather than by the order in
which ready threads are selected.  \tempo{} reuses the same deep-handler control
principle as the cooperative scheduler, but routes captured continuations
according to logical-time semantics rather than a single ready queue.

%% file: sections/language.tex
\tempo{} is a synchronous reactive library for \ocaml{} in direct style.  A
program is evaluated as a sequence of \emph{logical instants}.  During an
instant, computations propagate signal presence and values until no
same-instant work remains; at closure, signal presence is reset and delayed
continuations become eligible for the next instant.  The surface API is small,
and its types separate the three ingredients used by the rest of the paper:
signals, signal communication, and temporal control.

\paragraph{Signals and channel kinds.}
Signals are the only reactive communication medium.  \tempo{} distinguishes
event signals, which carry at most one value per instant, from aggregate
signals, which combine several emissions into one value:

\begin{lstlisting}[basicstyle=\ttfamily\footnotesize,frame=none]
type event
type aggregate
type ('emit, 'agg, 'mode) signal_core
type 'a signal = ('a, 'a, event) signal_core
type ('emit, 'agg) agg_signal =
  ('emit, 'agg, aggregate) signal_core

val new_signal : unit -> 'a signal
val new_signal_agg :
  initial:'agg -> combine:('agg -> 'emit -> 'agg) ->
  ('emit, 'agg) agg_signal
\end{lstlisting}

The generalized type \code{('emit,'agg,'mode) signal\_core} records this
distinction in the type of the channel.  The type \code{'emit} is the type of
values passed to \emitop{}; \code{'agg} is the type of values returned by
\awaitop{}; and \code{'mode}, through the phantom markers \code{event} and
\code{aggregate}, restricts the operations available on the signal.  For event
signals, emitted and observed values have the same type.  For aggregate
signals, emissions have type \code{'emit}, while awaiters observe the
accumulated value of type \code{'agg}.  The parameter \code{initial} supplies
the accumulator seed used for the first emission of each instant, and
\code{combine} updates the current aggregate with each emitted value.

\paragraph{Signal communication.}
Communication is performed by emitting on a signal and by waiting for a signal
value:

\begin{lstlisting}[basicstyle=\ttfamily\footnotesize,frame=none]
val emit : ('emit, 'agg, 'mode) signal_core -> 'emit -> unit
val await : ('emit, 'agg, 'mode) signal_core -> 'agg
val await_immediate : 'a signal -> 'a
\end{lstlisting}

Calling \emitop{} marks a signal as present in the current instant and
propagates a value.  On event signals, a second emission in the same instant is
rejected; on aggregate signals, each value is folded into the accumulator.
Waiting with \awaitop{} observes the signal value only after an instant
boundary, once the current instant has stabilized.  This boundary is essential
for aggregate signals, whose final value is known only at closure.  The
immediate form \awaitimmop{} is therefore restricted by its type to event
signals: when the awaited signal becomes present during the current instant, the
continuation can resume before the instant closes.

\paragraph{Temporal suspension and structured control.}
The remaining core primitives organize time and control:

\begin{lstlisting}[basicstyle=\ttfamily\footnotesize,frame=none]
val pause : unit -> unit
val when_ : ('emit, 'agg, 'mode) signal_core -> (unit -> unit) -> unit
val watch : ('emit, 'agg, 'mode) signal_core -> (unit -> unit) -> unit
val parallel : (unit -> unit) list -> unit
\end{lstlisting}

The operation \pauseop{} has no signal argument and explicitly defers the
current continuation to the next instant.  Presence guards are expressed with
\whenop{}: the body runs only when the guard signal is present in the current
instant, and nested guards act conjunctively.  Weak preemption is provided by
\watchop{}: if the watched signal becomes present during an instant, the scoped
body is not resumed after the boundary, while effects produced in that instant
remain visible.  Structured concurrency is expressed with \parallelop{}, which
launches several reactive branches and resumes the parent continuation only
after all branches have terminated.

\begin{lstlisting}[caption={Representative Tempo controller.},
label={lst:tempo-controller-compact},basicstyle=\ttfamily\footnotesize,frame=none]
let controller tick enabled stop out =
  watch stop (fun () ->
    let rec loop () =
      ignore (await_immediate tick);
      when_ enabled (fun () -> emit out ());
      pause ();
      loop ()
    in
    loop ())
\end{lstlisting}

Listing~\ref{lst:tempo-controller-compact} illustrates how the groups combine
in a controller.  The arguments \code{tick}, \code{enabled}, \code{stop}, and
\code{out} are signals.  In each instant where \code{tick} becomes present,
\awaitimmop{} lets the loop react before closure.  The guard \whenop{} makes
the emission of \code{out} depend on same-instant presence of \code{enabled},
and \pauseop{} separates two loop iterations by an instant boundary.  The
enclosing \watchop{} gives the loop a scoped stop condition: if \code{stop}
becomes present, current-instant effects are kept and the loop is not resumed
after the boundary.

%% file: sections/implementation.tex
\tempo{} is implemented as a small reactive machine delimited by an \ocaml{} 5
effect handler.  Surface operators perform typed requests; the scheduler turns
captured continuations into runtime tasks and assigns each task either to the
current instant, to a later instant, or to delayed work attached to a signal.
We use concrete \ocaml{} only to show this boundary, and present the rest of the
runtime abstractly.

\subsection{Machine State and Scheduling}

\paragraph{Machine state.}
At the abstract level, a scheduler state is \(S=\langle R,N,B,A\rangle\).
During an instant, \(R\) contains runnable tasks, \(N\) tasks delayed to the
next instant, \(B\) tasks whose \whenop{} guard set is not satisfied, and \(A\)
the active signal index.  Signals in \(A\) either have current-instant state or
stored frontiers, i.e. delayed control information attached to a signal.

\paragraph{Runtime records.}
The two main runtime objects use the following tuple convention:
\[
\begin{array}{rcll}
t & = & \langle \mathit{step}, \mathit{thread}, \mathit{guards},
          \mathit{kill}, \mathit{placement}\rangle
    & \text{runtime task},\\
s & = & \langle \mathit{present}, \mathit{value}, \mathit{awaiters},
          \mathit{guardWaiters}, \mathit{watchers}\rangle
    & \text{signal}.
\end{array}
\]
A runtime task is a scheduler-owned continuation of a logical behavior.  Its
\(\mathit{step}\) field is either an entry closure, such as the initial process
body or a body installed by \parallelop{}, \whenop{}, or \watchop{}, or a
captured continuation reified as \code{fun () -> continue k v}.  The other
fields record the join domain, active guards, active weak-preemption scopes,
and current owner of the task.  In the concrete record, ownership is implemented
by small local flags rather than as a single field.

A signal records current-instant observations and delayed control information.
The presence and value fields describe the instant.  The awaiter frontier stores
value waiters; the guard frontier stores blocked guarded tasks; the watcher
frontier stores weak-preemption tokens.  Event signals enforce one emission per
instant; aggregate signals accumulate values and deliver the final aggregate at
closure.  Concrete records also cache guard registrations and guard or kill
epochs for local hot-path checks.

\begin{lstlisting}[float=t,caption={Conceptual scheduler loop.},
label={lst:tempo-select-task},basicstyle=\ttfamily\footnotesize,frame=none]
procedure run_instant(S = <R,N,B,A>):
  while R is not empty:
    t <- pop_front(R)
    if not kills_alive(t):
      dispose(t)
    else if has_guards(t) and not guards_ok(t):
      block_on_missing_guards(S, t)
    else:
      handle_task(S, t)

  close_current_instant(S)

procedure run(S):
  while R is not empty or N is not empty:
    run_instant(S)
    promote_next_to_current(N, R)
\end{lstlisting}

\paragraph{Scheduler cycle.}
The engine alternates between an open phase and a closure phase.  The open phase
drains \(R\): dead kill contexts are discarded, guarded tasks whose dependencies
are missing move to \(B\), and the remaining tasks run under the handler.  The
protocols may enqueue work in \(R\) or \(N\), update frontiers, and mark signals
in \(A\).  Once \(R\) is empty, the instant is closed as described below, and
the next-instant queue becomes the next current queue.  The order used to drain
\(R\) is an implementation schedule; reactive results are determined by
logical-time propagation and instant closure.  Listing~\ref{lst:tempo-select-task}
abstracts away layout optimizations such as lazy pruning, sparse guard
deduplication, and guard/kill epoch caches, which preserve the same ownership
transfers.

\subsection{Effect Boundary and Protocols}

\paragraph{Typed requests.}
Listing~\ref{lst:tempo-effect-types} is the only \ocaml{} excerpt in this
section.  It gives the request type, two representative wrappers, and the
representative shape of the handler dispatch; the other surface operators use
the same one-line \code{perform} form, and the omitted handler branches use the
same \code{effect (...), k -> handle\_* ... k} delegation pattern.

\begin{lstlisting}[float=t,
caption={Effect requests, surface wrappers, and handler dispatch.},
label={lst:tempo-effect-types},
basicstyle=\ttfamily,
keywordstyle=\color{blue!60!black},
frame=none,
breaklines=true]
type _ Effect.t +=
  | New_signal : unit -> ('a, 'a, event) signal Effect.t
  | New_signal_agg : 'a * ('a -> 'e -> 'a) ->
      ('e, 'a, aggregate) signal Effect.t
  | Emit : ('e, 'a, 'm) signal * 'e -> unit Effect.t
  | Await : ('e, 'a, 'm) signal -> 'a Effect.t
  | Await_immediate : ('a, 'a, event) signal -> 'a Effect.t
  | Pause : unit Effect.t
  | Parallel : (unit -> unit) list -> unit Effect.t
  | When : ('e, 'a, 'm) signal * (unit -> unit) -> unit Effect.t
  | Watch : ('e, 'a, 'm) signal * (unit -> unit) -> unit Effect.t

let emit s v = perform (Emit (s, v))
let parallel ps = perform (Parallel ps)

let handle_task st t =
  match t.run () with
  | () -> handle_done ()
  | effect (Emit (s, v)), k -> handle_emit s v k
  | effect (Await s), k -> handle_await s k
  | effect Pause, k -> handle_pause k
  | effect (Parallel ps), k -> handle_parallel ps k
  | effect (When (s, body)), k -> handle_when s body k
  | effect (Watch (s, body)), k -> handle_watch s body k
  | ... (* other effects delegate to handle_* in the same way *)
\end{lstlisting}

\paragraph{Handler shape.}
The scheduler executes each task under one handler.  A normal return runs
\code{handle\_done}; an effect captures the continuation \(k\) and delegates to
the corresponding protocol.  The helpers close over the scheduler state and the
selected task, and therefore receive only the effect payload and \(k\).  Small
branches choose a destination for \(k\); \parallelop{}, \whenop{}, and
\watchop{} also install body tasks or join domains.  Each branch performs one
control transfer: continue now, schedule in \(R\) or \(N\), or store delayed
work in a signal frontier or join domain.

The following paragraphs cover the protocols named above and the omitted
branches; instant closure is described separately afterwards.  Placement in
\(R\) or \(N\) may use either stable or spawned ownership, as defined in the
final subsection.

\paragraph{Signal creation.}
\code{handle\_new\_signal} and \code{handle\_new\_signal\_agg} allocate a fresh
signal with empty await, guard, and watch frontiers, then execute
\code{continue k s}.  No task is moved between \(R\), \(N\), and \(B\).  The
signal is inserted into \(A\) only after an emission, await, or guard
registration makes it relevant to closure.

\paragraph{\emitop{}.}
\code{handle\_emit} updates the signal and resumes the emitter with
\code{continue k ()}.  The update marks \(s\) in \(A\).  Event signals set
presence and value, reject duplicate emissions, and consume their stored resume
plans according to \(\mathit{owner}\) and \(\mathit{dest}\).  Aggregate signals
fold the emitted value and delay awaiter delivery to closure, when the final
aggregate is known.  When an emission makes \(s\) present, guard waiters may
move from \(B\) to \(R\); watchers fire during closure.

\paragraph{\awaitop{} and \awaitimmop{}.}
The \code{Await} and \code{Await\_immediate} protocols differ only in the
destination chosen for a successful event resumption.  \awaitop{} always crosses
an instant boundary:
\[
\code{resume(v)} = \code{schedule\_next}(k,v).
\]
If an event signal is present when the request is handled, the plan is applied
at once but still places \(k\) in \(N\).  If the event is absent, or for
aggregate signals whose final value is known only at closure, the plan is stored
in \code{s.awaiters} and \(s\) is inserted into \(A\).  By contrast,
\awaitimmop{} is restricted to event signals and preserves same-instant
propagation: a present signal resumes inline, while an absent one stores the plan
\[
\code{resume(v)} = \code{schedule\_now}(k,v).
\]
Stable plans replace the retained task step by \(k\); spawned plans create a
fresh continuation task and mark the parent thread as suspended until resumption.
This separation makes the instant-boundary contract a scheduler property.

\paragraph{\pauseop{}.}
\code{handle\_pause} packages \(k\) as a continuation step, enqueues it in
\(N\), and returns without continuing inline.  The common unguarded path reuses
the current task; guarded pauses keep a separate continuation task so that guard
metadata remains attached across the instant boundary.  No signal frontier is
modified.

\paragraph{\parallelop{}.}
\code{Parallel(ps)} delegates to \code{handle\_parallel}.  For the empty list,
the handler continues \(k\) inline.  Otherwise, it creates one internal child
task per element of \code{ps}.  The children share a join domain, inherit the
parent's guard and kill metadata, and run in \(R\); the parent continuation is
registered as a stable or spawned join waiter.  Completion schedules the parent
continuation in \(R\) after all live children terminate.  Child creation is
internal: \parallelop{} is the effect request, while child tasks are runtime
objects.

\paragraph{\whenop{}.}
\code{When(s,body)} delegates to \code{handle\_when}.  The handler installs
\code{body} in \(R\) with the parent metadata extended by guard \(s\), reusing
the current task when possible.  Nested \whenop{} operators therefore become
conjunctive: the task must satisfy the whole guard list before it runs.  Missing
guards move the task to \(B\) and register it only on the corresponding
\code{guardWaiters} frontiers, using a direct path for one missing guard and the
sparse structure otherwise.  Emission returns the task to \(R\) only when all
missing guards are present and the kill context is alive.  If guards remain
absent until closure, the task is carried to \(N\).  Normal completion resumes
the parent continuation under the same handler boundary.

\paragraph{\watchop{}.}
\code{Watch(s,body)} delegates to \code{handle\_watch}.  If \(s\) is present,
the watched body is skipped and the parent continuation resumes inline: weak
preemption cannot start a region known to be killed.  Otherwise, the scheduler
installs \code{body} in \(R\) under a kill token associated with \(s\), reusing
the current task when possible, and registers a watcher.  If \(s\) is present at
closure, token cleanup schedules the parent continuation in \(N\); normal body
completion schedules it in \(R\).  A nested \watchop{} on a signal watched by an
ancestor is elided, and an absent watched signal is scanned only after emission
makes it part of \(A\).

\subsection{Instant Closure}

\begin{lstlisting}[float=t,caption={Conceptual instant-closure protocol.},
label={lst:tempo-close-instant},basicstyle=\ttfamily\footnotesize,frame=none]
procedure close_current_instant(S = <R,N,B,A>):
  for s in A:
    if present(s):
      fire_watchers(s.watchers)
      if aggregate(s):
        deliver_aggregate_awaiters(s.awaiters)
    else:
      prune_dead_frontiers(s)
    clear_presence_and_value(s)
    clear_current_guard_frontier(s.guardWaiters)
  A <- signals_with_persistent_frontiers(A)
  roll_blocked_tasks_to_next(B, N)
\end{lstlisting}

\paragraph{Closing an instant.}
When \(R\) is empty, closure scans \(A\).  Present signals fire watchers, and
present aggregate signals deliver the finalized aggregate to awaiters; event
awaiters have already been consumed by emission.  The scheduler clears transient
presence, values, and current-instant guard waiters, keeps only signals with
persistent frontiers, and rolls tasks still in \(B\) to \(N\).  This order makes
weak preemption non-retroactive: effects produced during the open phase remain
observable, while continuations under a dead kill context become ineligible from
the boundary onward.

\subsection{Ownership Discipline}

\paragraph{Stable task identity.}
The central implementation choice is stable task identity: a suspension does not
automatically destroy the selected task.  When sound, the scheduler replaces
its \(\mathit{step}\) field by the captured continuation; otherwise it stores
enough information to spawn a fresh continuation task later.  Signal awaiters
make this explicit with a resume plan
\[
\mathit{rp} =
  \langle \mathit{owner}, \mathit{dest}, k,
          \mathit{anchor}, \mathit{thread}, \mathit{guards}, \mathit{kill}\rangle,
\]
where \(\mathit{owner} \in \{\mathit{stable},\mathit{spawn}\}\) selects task reuse
or fresh-task creation, and
\(\mathit{dest} \in \{\mathit{now},\mathit{next}\}\) selects \(R\) or \(N\).  This
is the abstract counterpart of the four concrete signal-resumption cases in the
runtime.

\paragraph{Ownership transfers.}
The same distinction appears for \parallelop{} joins: stable waiters reuse a
retained parent task, while spawned waiters create a continuation task when the
children terminate.  Thus every handler branch is an ownership transfer:
continue inline, install the continuation in \(R\) or \(N\), move a guarded task
to \(B\), or store a resume plan in a signal frontier or join domain.  A signal
enters \(A\) when an emission, an \awaitop{}, or a guard registration makes its
frontiers relevant; a \watchop{} registration on an absent signal remains
dormant until emission makes the signal relevant to closure.

Across all protocols, handlers expose suspension points as continuations, and
the runtime assigns each continuation to the compartment required by logical
time.  The allocation discipline is local: a logical behavior keeps a stable
task identity whenever this does not change guard, join, or kill-scope
semantics.

%% file: sections/evaluation.tex
\label{sec:eval}

\paragraph{Questions and scope.}
The evaluation asks how much performance is lost when synchronous reactive
mechanisms are implemented as an \ocaml{} 5 library rather than by the
\reactiveml{} compiler and runtime.  We compare \tempo{} with two
\reactiveml{} executables: the legacy \reactiveml{} 1.09.07 runtime on
\ocaml{} 4.14.2, and a patched \reactiveml{} 1.09.07 build on \ocaml{} 5.4.1.
The comparison is therefore architectural rather than compiler-neutral:
\tempo{} and the patched \reactiveml{} share the \ocaml{} 5 toolchain, while
the legacy \reactiveml{} executable keeps the historical runtime environment.
The main suite isolates runtime mechanisms.  Each point is the median of 10
native runs; memory is peak RSS measured by the operating system.  The runs used
an Apple M5 Pro machine with 24 GB RAM and macOS 26.4.  Raw logs, metadata, and
plotting scripts are part of the artifact.  The mechanism benchmarks use
\(n\in\{10,20,40,80,160,320,640,1280,2560,5120,10240\}\); the supervision
scenario is measured up to \(n=1024\).  Before measurement, the regression suite
checks same-instant \awaitimmop{}, delayed \awaitop{}, \pauseop{} deferral,
\whenop{} guard scope, \watchop{} weak preemption, aggregate delivery, and
nested kill scopes.  The results therefore characterize the runtime of
Section~\ref{sec:implementation}, not a weakened semantics.

\paragraph{Workloads.}
The five benchmarks stress the main sources of scheduler work.  \bOne{} is the
multi-instant propagation-chain benchmark: a chain of \(n\) immediate awaits is
rebuilt and fired across
\(\max(2,\lfloor\log_2(\max(2,n))\rfloor)\) instants.  \bTwo{} broadcasts one
signal to \(n\) observers.  \bThree{} grows a binary parallel tree of depth
\(\lfloor\log_2(\max(2,n))\rfloor\).  \bFour{} is the multi-instant guarded
cascade benchmark: \(n\) nested guards are rebuilt and activated across the
same logarithmic number of instants as \bOne{}.  \bFive{} nests \(n\)
\watchop{} scopes and triggers weak preemption.  The multi-instant versions of
\bOne{} and \bFour{} are used directly in the main results, so there is no
separate validation subsection for them.

\begin{figure}[t]
\centering
\includegraphics[width=\textwidth]{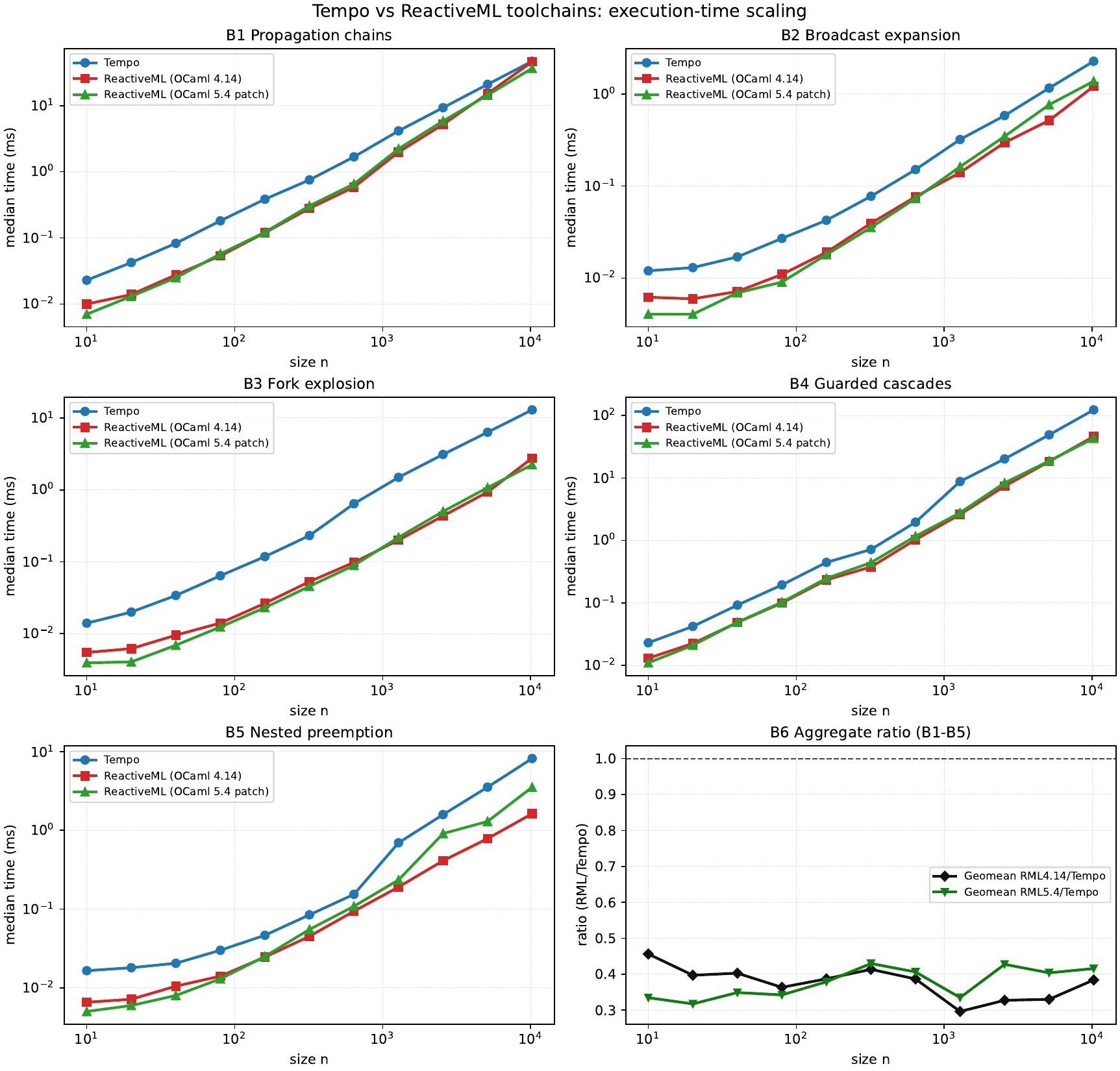}
\caption{Execution-time scaling for the five mechanism benchmarks.  B1 and B4
use the multi-instant variants.  B6 reports the geometric mean of
\(\mathrm{RML}/\tempo{}\) ratios over B1--B5.}
\label{fig:time-curves}
\end{figure}

\paragraph{Execution time.}
Figure~\ref{fig:time-curves} gives the scaling curves and the aggregate B6
ratio.  Ratios are \(\mathrm{RML}/\tempo{}\), so values below \(1\) indicate
that \tempo{} is slower.  The qualitative result is stable across the grid:
\tempo{} is generally slower than both \reactiveml{} variants, with only
isolated points near parity.  The geometric mean is \(0.327/0.427\) at
\(n=2560\) and \(0.330/0.404\) at \(n=5120\), for legacy/patched
\reactiveml{}.

The largest absolute times occur on the two multi-instant workloads, which
repeatedly create or update signals, continuations, and frontier entries.  At
\(n=2560\), \bOne{} takes \(9.316\) ms in \tempo{}, versus \(5.205\) ms and
\(5.856\) ms in legacy and patched \reactiveml{}; \bFour{} takes \(20.184\) ms,
versus \(7.385\) ms and \(8.339\) ms.  At \(n=5120\), the corresponding numbers
are \(20.971\) ms versus \(15.146/14.417\) ms for \bOne{}, and \(49.017\) ms
versus \(18.304/18.687\) ms for \bFour{}.  This is where the effect-based design
pays its largest control cost: each reactive suspension becomes an explicit
continuation managed by the scheduler.  \bThree{} shows the same effect through
task creation and join-domain management; \bTwo{} and \bFive{} show it through
broadcast wake-up and weak-preemption bookkeeping.

\begin{figure}[t]
\centering
\includegraphics[width=\textwidth]{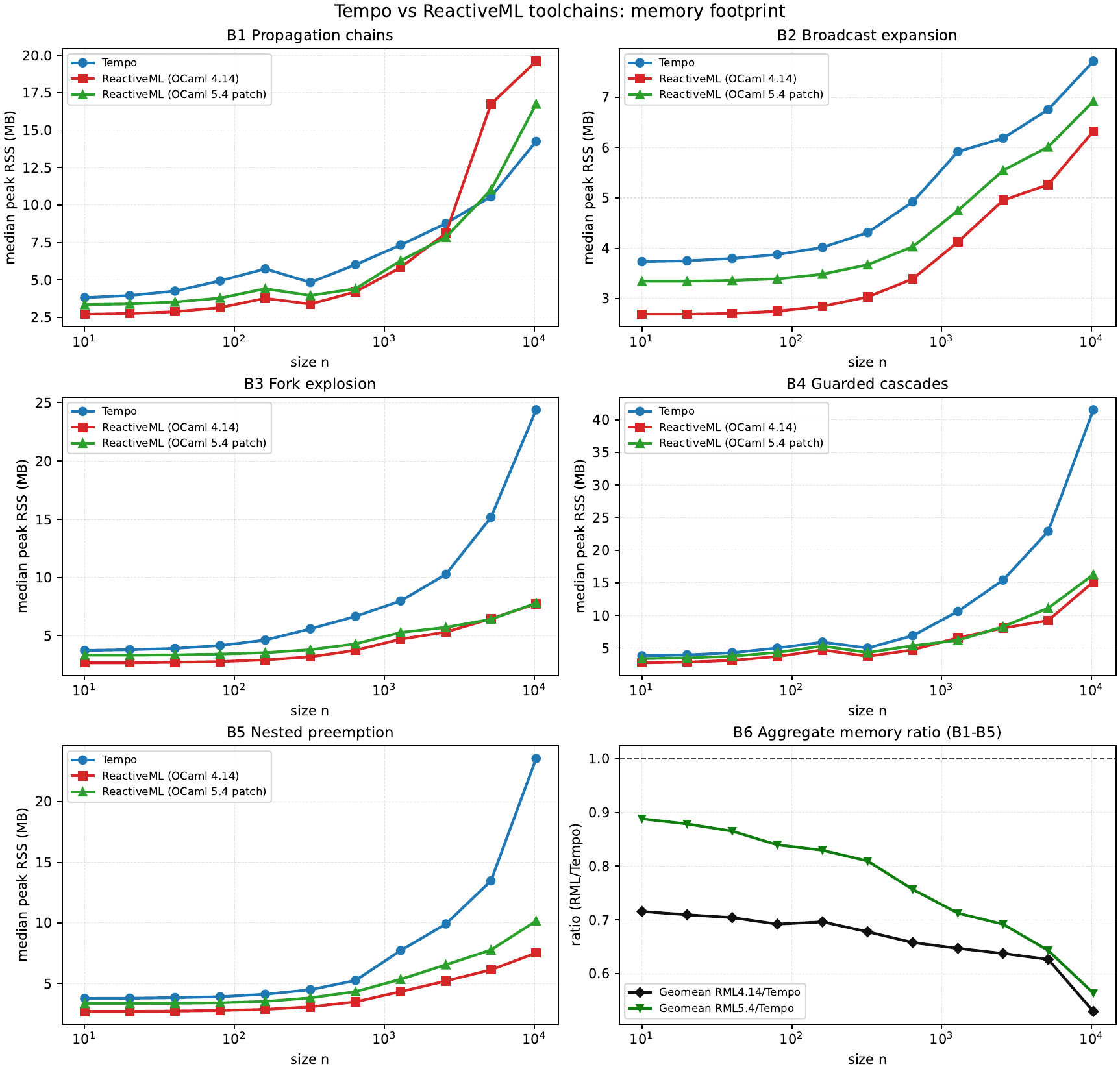}
\caption{Peak RSS for the same five benchmarks.  B6 reports the geometric mean
of \(\mathrm{RML}/\tempo{}\) memory ratios over B1--B5.}
\label{fig:memory-curves}
\end{figure}

\paragraph{Memory profile.}
Figure~\ref{fig:memory-curves} reports peak RSS on the same workload set.  The
aggregate \(\mathrm{RML}/\tempo{}\) memory ratio remains below \(1\), although
the propagation-chain benchmark is an exception: on that workload, the
\reactiveml{} executables have higher peak RSS in this campaign.  At
\(n=2560\), the aggregate memory ratio is \(0.638/0.692\); at \(n=5120\), it is
\(0.627/0.643\), for legacy/patched \reactiveml{}.

Except for the propagation-chain case, the gap follows from representation
rather than from a leak-like effect.  At \(n=2560\), \bOne{} uses \(8.766\) MB in
\tempo{}, versus \(8.109/7.844\) MB in legacy/patched \reactiveml{}; \bFour{}
uses \(15.422\) MB, versus \(8.062/8.281\) MB.  At \(n=5120\), \bOne{} uses
\(10.562\) MB in \tempo{} and \(16.750/11.016\) MB in \reactiveml{}, whereas
\bFour{} uses \(22.906\) MB and \(9.234/11.125\) MB.  \tempo{} materializes
tasks, captured continuations, awaiter links, guard registrations, kill-context
nodes, and watcher records as ordinary heap objects.  This explicit
representation increases bytes per live control edge; the \reactiveml{} compiler
can encode more of the control structure compactly.

\paragraph{Application-style scenario.}
The supervision scenario creates \(n\) sensor processes and runs them for a
small logarithmic number of rounds: 14 at \(n=128\), up to 20 at \(n=1024\).
Each round combines a presence guard, a tick, aggregate readings, a
cross-instant await, an observer notification, and a reset-driven weak
preemption of a maintenance process.  This synthetic, control-heavy composition
does not change the conclusion.  At \(n=1024\), \tempo{} takes \(15.698\) ms,
versus \(4.937/6.113\) ms in legacy/patched \reactiveml{}; peak RSS is
\(10.875\) MB, versus \(9.781/8.656\) MB.  The time ratios are \(0.315/0.389\).

\paragraph{Interpretation and threats.}
The central result is a cost profile, not a performance win for \tempo{}.  An
effect-based library can reproduce the reactive mechanisms of interest, but it
pays visible overhead for reifying continuations and scheduler metadata in the
host language, especially across many awaiters, repeated guard activation,
dynamic parallel structure, and preemption-heavy control.

The main threat to validity is benchmark scope: the mechanism workloads stress
isolated runtime features, and the supervision scenario remains a compact
stress case rather than a full application.  The toolchain comparison is also
asymmetric because legacy \reactiveml{} runs on \ocaml{} 4.14, although patched
\reactiveml{} provides a same-\ocaml{}-5 sensitivity check.  Finally, peak RSS
is a coarse process-level metric: it captures end-to-end memory pressure, but
not individual allocation sites.

%% file: sections/related.tex
\label{sec:related}

\tempo{} lies at the intersection of synchronous programming, reactive
execution frameworks, and continuation-based runtimes.  Its semantic lineage
includes Esterel~\cite{berry1992esterel},
Lustre~\cite{caspi1987lustre,halbwachs1991lustre},
Signal~\cite{benveniste2003synchronous}, and Boussinot's synchronous reactive
model with Reactive-C, Reactive Scripts, SugarCubes, and FairThreads
~\cite{boussinot2002reactive,boussinot1991reactivec,boussinot1996reactivescripts,boussinot1998sugarcubes,boussinot2006fairthreads}.
The closest language-level comparison is \reactiveml{}, which carries this model
into a strict, typed, higher-order functional language
~\cite{mandel2005reactiveml,mandel2006phd}; Lucid Synchrone is another reference
point for typed functional synchronous programming~\cite{pouzet2006lucid}.

The main distinction is where the reactive semantics is enforced.  Classical
synchronous systems and \reactiveml{} rely on dedicated language and compiler
machinery; \tempo{} reconstructs synchronous reactive execution as a library on
top of effect handlers.  Operationally, \reactiveml{} translates reactive
constructs to runtime combinators and continuation machinery, close to a
compiled control tree with dynamic instantiation and compact frontier metadata
~\cite{mandel2005reactiveml,mandel2006phd,mandel2015tenyears}.  \tempo{} keeps
this representation at library level: effects capture reactive steps, and one
scheduler manipulates tasks annotated with guard and kill metadata.

This use of effects connects \tempo{} to handler-based control abstractions
~\cite{plotkin2013effects,dolan2017multicore,dolan2020ocaml,leijen2017koka},
including the standard \code{spawn}/\code{yield} schedulers used in OCaml and
Multicore OCaml material~\cite{ocaml54effects,sivaramakrishnan2015effective},
and to continuation-oriented interpretations of control
~\cite{danvy1990control,filinski1994representing}.  Compared with generic
effect-based frameworks, however, \tempo{} fixes instant closure, delayed
absence, guard consistency, and weak preemption as mandatory semantics.  This
gives semantic clarity, but concentrates cost in control-heavy dependency
management.

%% file: sections/conclusion.tex
\label{sec:conclusion}

\tempo{} shows that core synchronous reactive mechanisms can be implemented at
library level in \ocaml{} with algebraic effects.  It remains slower than
\reactiveml{} on the tested campaign, especially with many awaiters, repeated
guard activation, dynamic parallelism, and weak preemption; the result is
feasibility and diagnosability, not performance parity.  Future work is twofold:
a correctness proof, ideally mechanized in Rocq or a related assistant, for
instant closure, delayed absence, guard activation, and weak preemption; and
\ocaml{}~5 domain-based parallel scheduling inside and across instants, a
direction unavailable to historical \reactiveml{} on \ocaml{} 4 except through a
separate runtime experiment
~\cite{dolan2017multicore,ocaml54effects,mandel2015tenyears}.